# Geometric spin dephasing of carriers with strong spin-orbit coupling.

Yuri A. Serebrennikov

Qubit Technology Center

2152 Merokee Dr., Merrick, NY 11566

ys455@columbia.edu

We will show that the geometric Berry-phase shift acquired by the components of the light hole (LH) spinor during an adiabatic collision is *equal* to the angular distance traveled by the wave vector. For LH, geometric dephasing leads to equality between angular and crystal momentum relaxation time. It will be shown that the same mechanism is completely ineffective for heavy and split-off holes in bulk crystals, and in $1P_{1/2}$ electron QSLs in spherical nanostructures. This non-model result suggests that these carriers may be considered as the stern candidates for implementation of dissipationless spin currents and fault-tolerant qubits.

72.25.Rb  03.65.Vf  03.67.Lx

In recent years, *all-electric* manipulation of spin motion in semiconductors has attracted a significant interest[1]. This interest stems largely from new opportunities for the hardware design of solid-state quantum computers where the spin rather than the charge of electron is used for information processing and storage. The purely electrical control of the spin in the absence of external magnetic fields may facilitate the integration of spintronics with traditional electronics. Clearly, electric field cannot directly affect the spin of a particle. Stark field may control the orbital momentum and, hence, through spin-orbit coupling (SOC), direct the spin. However, precisely because of SOC, the spin degrees of freedom are coupled with the nonmagnetic "bath" that destroys the entanglement in the spin subsystem, which is vital for quantum logical operations.

It is the common understanding that due to the strong SOC and the direct coupling of the crystal momentum $\vec{k}$ with the total angular momentum $\vec{J}$ the spin relaxation of holes in bulk crystals is on the scale of the momentum relaxation time, $\tau_k \sim 10^{-13}$ s. On the other hand, it has been found that the dephasing time of the electron spin in the excited quantum size levels (QSLs) in spherical CdSe nanocrystals is[2] $\sim 10^{-4}$ s. It has been shown that the admixture of the valence-band states near the surface of nanosphere leads to the spin-orbit splitting between the levels of the first excited QSL (1P) that is comparable with the fine-structure splitting in alkali atoms (~ 1 - 10 meV for quantum dot radius 40 - 13 Å)[2]. Thus, the question arises why such a strong SOC does not guarantee ultra fast spin relaxation.

50 years ago, R.J. Elliott[3] introduced the spin-orbit interaction into the band theory of semiconductors. It has been shown that presence of SOC in the crystal Hamiltonian leads to a mixing of "spin-up" and "spin-down" electron Bloch functions in the different conduction bands with the same $\vec{k}$. Due to this mixing the *adiabatic* Coulomb scattering of a wave vector $\vec{k} \to \vec{k}'$ results in a nonzero spin-flip probability $P \sim \sin^2 \vartheta_{\vec{k},\vec{k}'}$, where $\vartheta_{\vec{k},\vec{k}'}$ is the angle between $\vec{k}$ and $\vec{k}'$. The theory was limited by the constrain $\lambda / \Delta E \ll 1$, where $\Delta E$ is the interband energy separation and $\lambda$ characterize the amplitude of the interband matrix element of SOC. This condition is well satisfied for *s*-like electrons in the lowest conduction-band, but clearly is not applicable, e.g., for near degenerate valence bands in ordinary semiconductors where the intrinsic SOC caused by the atomic cores is rather strong (296 mEv in Ge).

Very recently[4], it has been shown that Elliott's spin-flip mechanism arises from the random acquisition of geometric phases and provides the *intrinsic* sink of spin coherence in the system. Due to weak SOC of s-like electrons in the lowest conduction band, this "geometric dephasing" yields relatively small spin relaxation rate described by the Elliott relationship [] $1/T_s \sim (\lambda/\Delta E)^2 \tau_k^{-1}$. Although, the method of the theory is rather general and is applicable for systems with arbitrary strong SOC it does not provide a clear physical insight into the problem.

Here we will show that the geometric Berry-phase shift acquired by the components of the LH spinor during an adiabatic collision is *equal* to the angular distance, $\vartheta_{\vec{k},\vec{k}'}$, traveled by $\vec{k}$. The corresponding Bloch vector (effective spin) *precisely* follows the rotation of the wave vector. Consequently, stochastic scattering of the LH momentum leads to $T_s = \tau_k$. Remarkably, the same mechanism is completely ineffective for heavy (HH) and split-off (SO) holes in bulk crystals, and in $1P_{1/2}$ electron QSLs in spherical nanostructures, where relaxation may occur only through *nonadiabatic* interband transitions. This non-model result suggests that these carriers may be considered as the stern candidates for implementation of dissipationless spin currents[5] and fault-tolerant qubits[6,7].

In the Elliott mechanism of spin relaxation the loss of coherence occurs only in the short time intervals *during* collisions. To describe the evolution of the system throughout the particular scattering event it is convenient to transform the basis into the moving (*M*) frame of reference that follows the rotation of $\vec{k}$, $\Psi^{(M)}(t) = R(t)\Psi^{(L)}(t)$, where $\Psi$ is the instantaneous eigenvector of the total, *nontruncated* Hamiltonian of the crystal, *H*. The rotation operator $R = \exp[-i\phi\hat{n}\vec{J}]$ maps the space-fixed lab (*L*) frame into the actual orientation of the *M*-frame at time *t*. Recall that in the presence of SOC, the Bloch functions are not factorizable into the orbital and spin parts, hence, $\vec{J} = \vec{L} + \vec{S}$, is included into the transformation $R(t)$. The quantization axis of the system is conveniently chosen along the direction of $\vec{k}$ and corresponds to the $z_M$ axis of the *M*-frame. Then the unit vector $\hat{n}$ denotes the instantaneous axis of the $\vec{k} \to \vec{k}'$ rotation and the angle of this rotation is denoted as $\phi$. Note that in the *M*-frame, the total Hamiltonian of the problem must be replaced with

$$\tilde{H}^{(M)}(t) = H^{(M)}(0) - iR^{-1}\dot{R} = R^{-1}H^{(L)}(t)R - iR^{-1}\dot{R}. \qquad (1)$$

The first term, $H^{(M)}(0)$, is static, while the second one is the trivial gauge potential that appears in the *M*-frame. Up to this point, the transformation is exact. The standard next step that leads to a nontrivial gauge potential entails adiabaticity of the $\vec{k} \to \vec{k}'$ rotation. Physically this mean that the inverse dwell time of the collision, $\tau_c$, is much smaller than the characteristic frequency inside the scattering particle, i.e., it requires vanishing interband transitions during the scattering event, $\Delta E \gg 1/\tau_c$. Since $\tau_c \cong \tau_k$, this condition is well satisfied for *s*-like electrons in the lowest conduction band. However, this may not be the case for near degenerate bands, e.g., holes close to the center of the Brillouin zone. Thus, to proceed further we need a thorough understanding of the valence bands Hamiltonian.

In bulk semiconductors such as Si, Ge or GaAs in the absence of spin the top of the valence band at the Γ point ($\vec{k} = 0$) consist of three degenerate *p*-like functions that within the "spherical approximation"[8][9] transform under the full rotation group as eigenfunctions of effective orbital momentum $L = 1$. Near the center of the Brillouin zone the second order (effective-mass) $\vec{k} \cdot \vec{p}$ perturbation theory gives a Hamiltonian matrix every element of which is a quadratic function of the components of $\vec{k}$. Therefore, without spin a rotationally invariant effective $\vec{k} \cdot \vec{p}$ Hamiltonian that acts in the space of *p*-functions contains only two terms proportional to $k^2 = \sum k_i^2$ and $(\vec{k} \cdot \vec{L})^2$. In the Cartesian basis ($i = x, y, z$) the latter term can be expanded in the following form[10]

$$(\vec{k} \cdot \vec{L})^2 = k^2[(2/3)I + \sum_{i,j} \hat{k}_i \vec{\vec{Q}}_{ij} \hat{k}_j] \qquad (2)$$

Here $\vec{\vec{Q}}_{ij} = (1/2)[L_i L_j + L_j L_i - (4/3)\delta_{ij} I]$ is the traceless symmetric "quadrupole" tensor, $\hat{k}_i$ is the component of the unit vector $\vec{k}/|\vec{k}|$, and $I$ is the identity matrix ($3 \times 3$ in this case). Tensor $\vec{\vec{Q}}_{ij}$ can be expressed in terms of the components of a second rank irreducible orthonormal tensor operator of the full rotation group $T_{2q}(L) = \sum_{\mu\mu_1} C^{2q}_{1\mu 1\mu_1} L_\mu L_{\mu_1}$, where $C^{2q}_{1\mu 1\mu_1}$ is the Clebsch-Gordon coefficient. In the *principal-axes* system (*M-frame*) of the quadrupole tensor: $\vec{\vec{Q}}_{z_M z_M} = (2/3)^{1/2} T_{20}(1)$

$\vec{\vec{Q}}_{x_M x_M} = [T_{22}(1) + T_{2-2}(1)]/2 - (6)^{-1/2} T_{20}(1)$, $\vec{\vec{Q}}_{y_M y_M} = -[T_{22}(1) + T_{2-2}(1)]/2 - (6)^{-1/2} T_{20}(1)$,

and it is straightforward to obtain the following expansion of the kinetic *spinless* Hamiltonian ($L = 1$, $S = 0$)

$$H^{(M)}_{k^2}(0) = (a/2m_0)(k^2/3)I + (2/3)^{1/2} D_k T^{(M)}_{20}(1) + E_k[T^{(M)}_{22}(1) + T^{(M)}_{2-2}(1)], \qquad (3)$$

where we introduced the notation

$$(m_0/b)D_k = -(2k_{z_M}^2 - k_{x_M}^2 - k_{y_M}^2)/2, \quad (m_0/c)E_k = -(k_{x_M}^2 - k_{y_M}^2)/2,$$

$m_0$ is the bare electron mass, $\hbar = 1$, and the coefficients $a$, $b$, and $c$ are the dimensionless parameters of the theory. Expansion (3) can be written in the more compact form as

$$H_{k^2}^{(M)} = (K/3)I + \sum_q (-1)^q K_{2q}^{(M)} T_{2-q}^{(M)}(1), \qquad (4)$$

where $K_{20}^{(M)} = (2/3)^{1/2} D_k$, $K_{2\pm 1}^{(M)} = 0$, $K_{2\pm 2}^{(M)} = E_k$. The first term on the RHS of Eq.(4) represents an *isotropic* part of the kinetic energy of the carrier, $K = TrH_{k^2}$. The second term, a scalar product of two spherical tensors of rank 2, is clearly *anisotropic*, $[L^2, H_{k^2}] \neq 0$. Physically this means that that the kinetic motion of the hole breaks the isotropy of the system. Outside the center of the Brillouin zone, the triplet of degenerate valence bands will split apart, $L^2$ and $\vec{L}$ are no longer a constants of motion. In fact, with the substitution $\vec{E} \rightarrow \vec{k}$ in Eq.(4), careful readers will recognize in $H_{E^2}$ the familiar form of the quadratic crystal field Hamiltonian that may completely lift the degeneracy and "quench" the orbital momentum. We will return to this point later. For axially symmetric systems ($E_k = 0$) $L_{z_M}$ does commute with $H_{k^2}^{(M)}$ and Eq.(4) yields eigenstates that can be characterized by the helicity: $m_L = \hat{\vec{k}} \cdot \vec{L}^{(M)}$; $m_L = \pm 1, m_L = 0$ with the corresponding eigenvalues $E_\pm = K/3 + D_k/3$ and $E_0 = K/3 - 2D_k/3$.

In the presence of spin the combined action of *isotropic* spin-orbit interaction $H_{SO} = \lambda \vec{L} \cdot \vec{S}$, where $\lambda$ is the parameter of spin-orbit splitting at $\vec{k} = 0$, and *anisotropic* $H_{k^2}$ will split the sixfold degenerate manifold of the valence bands into a series of states that for a given value of $\vec{k}$ can be described by the spinor envelope functions of the compound L-S system: $|LS, Jm; \vec{k}> = \sum_{\mu \mu_1} C^{Jm}_{L\mu\, 1/2\mu_1} |L\mu; \vec{k}> |1/2\mu_1>$.

The matrix elements of $H_{SO}$ and $T_{kq}(L)$ in the $|LS, Jm>$ basis are well known, which allows us to represent the $6 \times 6$ matrix of the $\vec{k} \cdot \vec{p}$ Hamiltonian, $H^{(M)}(0) = H_{SO} + H^{(M)}_{k^2}$, in the following form

$$<1\,1/2, J_1 m_1; \vec{k}|H^{(M)}|1\,1/2, Jm; \vec{k}> = \frac{\lambda}{2}[J(J+1) - \frac{11}{4}]\delta_{JJ_1}\delta_{mm_1} + K/6\,\delta_{JJ_1}\delta_{mm_1}$$
$$+ (-1)^{J+3/2}(2J+1)^{1/2}(5)^{1/2}\begin{Bmatrix} 2 & 1 & 1 \\ 1/2 & J_1 & J \end{Bmatrix}[\sqrt{\frac{2}{3}}D_k C^{J_1 m_1}_{Jm\,20} + E_k(C^{J_1 m_1}_{Jm\,22} + C^{J_1 m_1}_{Jm\,2-2})] \quad (5)$$

As expected, Eq.(5) has the same analytical structure as Luttinger-Kohn Hamiltonian[11] (*in the M-frame*) and apparently reproduces the form of Luttinger Hamiltonian ($J = J_1 = 3/2$, $4 \times 4$ block). Obviously, $[J^2, H^{(M)}] \neq 0$, however, if the anisotropic part of $H^{(M)}_{k^2}$ is axially symmetric, $J_{z_M}$ is conserved, and the eigenfunctions of $H^{(M)}$ can be classified by the helicity $m = \hat{\vec{k}} \cdot \vec{J}^{(M)}$. Bands with $J = 3/2, m = \pm 3/2$ correspond to HH; with $J = 3/2, m = \pm 1/2$ to LH; bands with $J = 1/2, m = \pm 1/2$ represent the SO holes. Due to *T*-invariance of the model (no magnetic interactions) each of these bands has Kramers degeneracy.

Expression (5) tells us that at $\Gamma$ point, similar to a fine structure splitting in isolated atoms, SOC, being an intrinsic part of the crystal Hamiltonian, breaks up the six-fold valence band degeneracy into multiplets of $\vec{J}$ (splitting $= 3\lambda/2$), but preserves the isotropy of the system. Anisotropy comes from the kinetic motion of the hole that, similar to a crystal field, is responsible for further lifting of the degeneracy of the $\Gamma_8$ states into HH and LH bands. As already mentioned, the *intrinsic* SOC is rather strong in common semiconductors, hence, $3\lambda/2 \gg 1/\tau_c$, and it is safe to ignore nonadiabatic transitions between $J = 3/2$ and $J = 1/2$ bands.

Taking onto account Eqs.(1) and (5) it can be easily shown that transformation into the rotating *M*-basis of the $J = 3/2$ multiplet that is *adiabatically isolated* from the rest of the band structure yields

$$\widetilde{H}^{(M)}(t) = (K/4)I + (2/3)^{1/2}\sum_{q}(-1)^q K_{2q}^{(M)} T_{2-q}^{(M)}(J) + \vec{\omega}(t)\cdot\vec{J}^{(M)}. \qquad (6)$$

Here $\vec{\omega}(t)$ is an instantaneous angular velocity of the *M*-frame relative to the *L*-frame at time *t*: $\omega_{x_M} dt = -\sin\theta\, d\varphi$, $\omega_{y_M} = d\theta$, $\omega_{z_M} dt = \cos\theta\, d\varphi$; spherical angles, $\theta$ and $\varphi$, define the orientation $\hat{\vec{n}}(t)$, and we replace $T_{2q}(L=1)$ with the operator equivalent

$$T_{2q}(J = 3/2) = (5/4)^{1/2}\sum_{m m_1} C_{Jm\,2q}^{Jm_1} |Jm_1\rangle\langle Jm|.$$

The last term on the RHS of Eq.(6) represents nontrivial (generally) nonadiabatic and non-Abelian gauge potential that appears in the rotating frame. Remarkably, it has the form of a Coriolis interaction and may be interpreted as "fictitious magnetic field" acting along the respective rotation axis. Eq.(6) depends on a choice of gauge that specifies the reference orientation, i.e. the orientation in which the *M*-frame coincides with some space-fixed frame. At the moment $t = 0$, this orientation may always be chosen such that the axis $z_M$ represents the quantization axis of $\vec{J}^{(M)}$. Noticeably, with the obvious redefinition of the parameters, $\widetilde{H}^{(M)}(t)$, Eq.(6), coincides with the expression for the effective *nuclear* quadrupole Hamiltonian of spin $I = 3/2$ obtained in the rotating frame[12].

In the axially symmetric case ($E_k = 0$), the matrix of $H^{(M)}(0)$ is diagonal with the eigenvalues $E_{HH} = K/4 + D_k/3$, $E_{LH} = K/4 - D_k/3$. Now if we neglect the nonadiabatic transitions between HH and LH bands, $|D_k| \gg |\vec{\omega}|$, we obtain two adiabatically isolated Kramers doublets: $\widetilde{H}_{HH}^{(M)}(t) = (3/2)\sigma_{z_M}^{(M)}\omega_{z_M}$ and $\widetilde{H}_{LH}^{(M)}(t) = (1/2)[\sigma_{z_M}^{(M)}\omega_{z_M} + 2(\sigma_{x_M}^{(M)}\omega_{x_M} + \sigma_{y_M}^{(M)}\omega_{y_M})]$, where $\vec{\sigma}$ is the vector of Pauli matrices. Clearly, the non-Abelian structure is only present in the LH band. However, if we suppose that during a short time of a collision $|\vec{\omega}|$ and the plain of the $\vec{k} \to \vec{k}'$ rotation remain constant than the axis of rotation $\hat{\vec{n}}$ can be assigned to $x_M$, $\theta = \pi/2$, and the LH gauge potential lost its non-Abelian character, $\widetilde{H}_{LH}^{(M)}(t) = \sigma_{x_M}^{(M)}\omega_{x_M}$. Notably, this leads to $\widetilde{H}_{HH}^{(M)}(t) = 0$.

To describe the evolution of the observable Bloch vector that represents Kramers doublet during a collision in the *local* reference (*space fixed*) frame we have to perform a reverse rotation of the basis compensating for the rotation of the M-frame (see Ref.[4] for details), which for LH yields the following result:

$$\vec{u}_Z^{(L)}(t) = Tr[\vec{\sigma}_Z^{(L)} \exp(i\omega_{x_M}\vec{\sigma}_{x_M}^{(M)} t/2) \vec{\sigma}_Z^{(L)}] = \cos(\omega_{x_M} t) = \cos\vartheta_{\vec{k},\vec{k}'}. \qquad (7)$$

We would like to emphasize that this expression is only applicable during the collision ($t \leq \tau_c, \omega t \leq \pi$), in the *L*-frame that reflects the geometry of the particular scattering event. Nevertheless, it clearly shows that the geometric Berry-phase shift acquired by the components of the LH spinor during an adiabatic collision is *equal* to the angular distance traveled by $\vec{k}$. Increasing linearly in time, this phase shift is equivalent to spin-precession or zero-field splitting. The corresponding LH Bloch vector *precisely* follows the rotation of the wave vector. Consequently, stochastic scattering of the LH momentum leads to $T_s = \tau_k$. Thus, geometric dephasing is the main source of the angular momentum relaxation in LH bands. On the other hand, because the gauge field only connects states with helicity difference $\Delta m = 0, \pm 1$, this mechanism is ineffective in the HH bands. Within the adiabatic limit, the effective spin in HH bands is conserved, relaxation may occur only due to *nonadiabatic* interband transitions, i.e. with a much smaller rate that strongly depends on the ratio between the HH-LH splitting $|D_k|$ and $|\vec{\omega}|$. This important non-model result may qualitatively explain the dramatic increase in the HH spin lifetime, $T_s \sim 1$ ns, observed in quantum wells[13]. It is easy to check that due to geometry of the collision the effective Hamiltonian of the adiabatically isolated SO band $\widetilde{H}_{SO}^{(M)}(t) = 0$. Since SO bands are fully populated this conclusion, however, has no practical merit.

The advantage of the expansion of the $\vec{k} \cdot \vec{p}$ Hamiltonian in terms of irreducible tensor operators, Eq.(), is that it can be easily generalized to the case where external electric field, strain or the confinement potential is present. As remarked earlier, the substitution $\vec{E} \to \vec{k}$ leads to the quadratic crystal field Hamiltonian that may suppress SOC in the system by making $D_E$ comparable with or larger than $\lambda$. The similar effect one may expect in highly porous silicon or anisotropic nano- and hetero-structures, where the confinement potential lifts the HH-LH degeneracy at $\Gamma$ point. The HH-LH splitting in typical 2D semiconductor nanostructures[14] $\geq 50$ mEv at $\vec{k} = 0$. This is large enough to react adiabatically to any perturbation caused by the rotation of $\vec{k}$ near the zone center. From this perspective, it becomes clear why the hole's spin relaxation time was reported to be a rapidly decreasing function of the HH population temperature in the GaAs quantum-wells, where merely the HH spin quantum beats were observed[15].

Finally, isomorphism between $H_k^{(M)}$ and $H_E^{(M)}$ Hamiltonians allows to apply the obtained results to 1P excited electron QSLs in spherical nanostructures, where one may expect infinite spin lifetime in the $1P_{1/2}$ state, provided the adiabatic condition is fulfilled.

[1] Y. Kato, R. C. Myers, A. C. Gossard, and D. D. Awschalom, Nature **427**, 6969 (2004); Y. Kato, R. C. Myers, D. C. Driscoll, A. C. Gossard, J. Levy, and D. D. Awschalom, Science **299**, 1201 (2003); J. B. Miller, D. M. Zumbuhl, C. M. Marcus, Y. B. Lyanda-Geller, D. Goldhaber-Gordon, K. Campman, and A. C. Gossard, Phys. Rev. Lett. **90**, 076807 (2003).

[2] A. V. Rodina, Al. L. Efros, and A. Yu. Alekseev, Phys. Rev. B **67**, 155312 (2003).

[3] R. J. Elliott, Phys. Rev. **96**, 266 (1954).

[4] Yu. A. Serebrennikov, Phys. Rev. Lett., Accepted for publication (2004).

[5] S. Murakami, N. Nagaosa, and S. C. Zhang, Science **301**, 1348 (2003).

[6] Yu. A. Serebrennikov, Phys. Rev. B. **70**, 064422 (2004).

[7] B. A. Bernevig and S.-C. Zhang, arXiv: quant-ph/0402165 (2004).

[8] J. M. Luttinger, Phys. Rev. **102**, 1030 (1956).

[9] N. O. Lipari and A. Baldereschi, Phys. Rev. Lett. **25**, 1660 (1970).

[10] D. A. Varshalovich, A. N. Moskalev, and V. K. Khersonsky, *Quantum Theory of Angular Momentum*, (World Scientific, Singapore, 1989)

[11] J. M. Luttinger and W. Kohn, Phys. Rev. **97**, 869 (1955).

[12] R. Tycko, Phys. Rev. Lett. **58**, 2281 (1987); J. W. Zwanziger, M. Koenig, and A. Pines, Phys. Rev. A. **42**, 3107 (1990); S. Appelt, G. Wackerle, and M. Mehring, Phys. Rev. Lett. **72**, 3921 (1994).

[13] Ph. Roussignol, P. Rolland, R. Ferreira, C. Delande, and G. Bastard, Phys. Rev. B **46**, 7292 (1992).

[14] G. Sun, L. Friedman, and R. A. Soref, Phys. Rev. B **62**, 8114 (2000).

[15] X. Marie, T. Amand, P. Le Jeune, and P. Renucci, Phys. Rev. B **60**, 5811 (1999).